\documentclass[useAMS,usenatbib]{mn2e}

\usepackage{graphicx}

\title{Extreme luminosities in ejecta produced by intermittent outflows around rotating black holes} 
\author[M.H.P.M. van Putten]
{Maurice H.P.M. van Putten$^1$ \thanks{E-mail: mvp@sejong.ac.kr}\\
\mbox{}\\
$^1$ Astronomy and Space Science, Sejong University, 98 Gunja-Dong Gwangin-gu, Seoul 143-747, Korea}
\begin{document}

\date{}

\pagerange{\pageref{firstpage}--\pageref{lastpage}} \pubyear{2002}

\maketitle

\label{firstpage}

\begin{abstract}
Extreme sources in the Transient Universe show evidence of relativistic outflows from intermittent inner engines, such as cosmological gamma-ray bursts. They probably derive from rotating back holes interacting with surrounding matter. We show that these interactions are enhanced inversely proportional to the duty cycle in advection of magnetic flux, as may apply at high accretion rates. We demonstrate the morphology and ballistic propagation of relativistic ejecta from burst outflows by numerical simulations in relativistic magnetohydrodynamics. Applied to stellar mass black holes in core-collapse of massive stars, it provides a robust explosion mechanism as a function of total energy output. At breakout, these ejecta may produce a low-luminosity GRB. A long GRB may ensue from an additional ultra-relativistic baryon-poor inner jet from a sufficiently long-lived intermittent inner engine. The simulations demonstrate a complex geometry in mergers of successive ejecta, whose mixing and shocks provide a pathway to broadband high energy emission from magnetic reconnection and shocks. 
\end{abstract}


\section{Introduction}

The Transient Universe reveals extreme relativistic sources with evidence for inner engines harboring black holes.
Foremost examples are cosmological gamma-ray bursts (GRB) associated with stellar mass black holes, the majority of which are long duration events believed to originate in aspherical supernovae  \citep{pap89,hof99} within the larger class of core-collapse supernovae (CC-SNe, \cite{mau10}). It points to a jet-powered explosion mechanism \citep{mac99} as a relativistic variant to a magnetic wind powered explosion from an angular momentum rich inner engine \citep{bis70}. While CC-SNe are fairly common, it should be mentioned that the branching ratio into LGRBs is quite small, at most a few percent of the Type Ib/c supernovae \citep{van04,gue07}. The mechanism for producing CC-SNe is therefore quite robust compared to producing a successful GRB. It is widely believed that a closely related mechanism produces relativistic outflows in their supermassive counterparts in active galactic nuclei (AGN).
 
Astrophysical black hole are generally rotating. According to the Kerr metric, they rotate fast than matter at the Inner Most Stable Circular Orbit (ISCO) whenever the energy in angular momentum exceeds a mere 5.3\% of their maximal spin energy. Black holes have a remarkably broad window, therefore, for energetic interactions with and even {\em onto} surrounding matter via an inner torus magnetosphere \citep{van99a}. We expect a turbulent inner disk by competing torques from magnetic surface stresses, i.e., by feedback from the black hole and outgoing winds, shaping it into a torus by induced thermal and magnetic pressure.

Extremely luminous GRBs further show burst-like behavior in outflows from a possibly intermittent inner engines, providing the conditions
recurrent shocks in relativistic outflows in the internal shock model of the prompt gamma-ray emission in GRBs \citep{ree94,pir97,lev97,fen99,kob97,dai98}.

Long GRBs from the Burst and Transient Source Experiment (BATSE, \cite{kou93}) show a pronounced positive correlation between luminosity in variability \citep{ste99,rei01}. Possibly related is the more recently discovered {\em Swift} class of short GRBs with Extended Emission (SGRBEE) and long GRBs with no association to supernovae (LGRBN). Their long/soft tails share the same Amati correlation as normal LGRBs but with relatively small isotropically equivalent energies ($E_{iso}$), perhaps as a result of their different astronomical origin \citep{van14}. Micro-quasars such as GRS 1915+105 \citep{mir94} and quasars such as 3C273 and 3C279 \citep{pea81,weh01} are well known for their extreme luminosities and blobs appearing at superluminal velocities (e.g. \cite{ant13}). Conceivably, they share a similar emission mechanism, whose luminosity is somehow enhanced by intermittency in the inner engine. In particular, the interaction of a black hole onto matter at the ISCO may be inherently unstable or unsteady, e.g., by superradiant scattering \citep{van99}, hydrodynamic or magnetic-pressure induced instabilities \citep{van03}, accretion from an inhomogeneous disk \citep{dex11} or otherwise. These intermittencies will result in burst outflows in magnetic winds from an inner disk or torus winds, possibly  along with burst outflows in ultra-relativistic jets along the black hole spin axis. 

Here, we ask: Is luminosity enhanced by intermittency, circumventing bounds that may exist in the more restricted class of steady state models? To this end, we consider the luminosity and morphology of burst-like outflows in magnetic disk or torus winds as a function of duty cycle, defined by the ratio of the duration $\tau$ of the on-state to a (quasi-)period $T$ of the on- and off-state, and accretion rate, as a boundary condition to an extended accretion disk. 

The luminosity of magnetic winds derives from poloidal magnetic flux by Faraday induction, whereas their total energy output derives from accretion and, possibly, the spin energy of a central black hole. Hyper-accretion in core-collapse of massive stars \citep{woo86} may advect magnetic flux, leading to build up of potentially strong magnetic flux around the black hole as a function of accretion rate (cf. \cite{kum08b}). 
Alternatively, feedback from the black hole may power a dynamo in matter about the ISCO. Strong magnetic fields may develop even in the absence of accretion, e.g., in mergers of neutron stars with another neutron star or black hole companion. Either way, the energy in the poloidal magnetic field is subject to a stability limit in quasi-stationary models \citep{van03}. 

In what follows, we use the theory of relativistic ideal magnetohydrodynamics (RMHD). RMHD is well suited for studying outflows from compact sources by their large hydrodynamical and magnetic Reynolds numbers. It contains fast and slow magnetosonic waves as well as Alfv\'en waves. For rotating systems, Alfv\'en waves are crucial to shedding angular momentum in magnetic winds along helical magnetic fields that may extend, if open, out to infinity. Exposed to a poloidal magnetic field from an inner accretion disk, a rotating black hole develops an equilibrium magnetic moment that may similarly support open magnetic field lines out to infinity. Subject to frame dragging, measured around the Earth \citep{ciu04,eve01}, blobs of charged particles can be ejected in their radiative Landau states by a powerful spin-orbit coupling resulting from Papapetrou forces \citep{pap51,van05}, likewise shedding angular momentum to infinity.

Crucial to the formation of magnetized outflows are the conditions at launch, and mostly so at the inner disk down to the ISCO
where the disk luminosity in winds is largest \citep{lov02,lov03}. 
Away from shocks, ideal MHD describes the adiabatic evolution of a magnetized fluid. Non-equilibrium pressure conditions at launch hereby tend to propagate along with the outflow downstream, allowing distinctive complex time-dependent jet morphology to develop throughout the life of the jet. In addition to the distinctive nose-cone of toroidally magnetized jets marking the head of the jet \citep{cla86,laz12}, these conditions
can produce magnetized toroidal vorticity (MTV, in the shape of a shoulder) propagating at intermediate velocity \citep{van96}. The MTV-nose cone morphology is a sharp departure from hydrodynamical jets, where a hydrodynamical toroidal vorticity (HTV) marks the head of the jet. It represents back flow forming a cocoon of shocked jet flow its wake, that envelopes the unshocked jet. 

We now argue that intermittencies at the source of outflows are advantageous to the launch of powerful winds. In transverse MHD, poloidal flux is advected in proportion to mass accretion rate, $\dot{m}(t)$. Since associated instantaneous luminosity $L_w(t)$ in magnetic winds scales with the the square of poloidal magnetic flux, we have $L_w(t)\propto \dot{m}^2(t)$. Given a duty-cycle defined by a typical duration $\tau$ of bursts repeated over intervals of order $T$, we obtain \citep{van99}
\begin{eqnarray}
<L_w^{i}> = \frac{4}{3} L_w^c\left(\frac{T}{\tau}\right),
\label{EQN_LL}
\end{eqnarray}
where the superscript $i$ refers to the intermittent luminosity and $c$ refers to the equivalent continuous luminosity 
at a given mean mass accretion rate $\dot{m}_0$ far out in the extended accretion disk. 

The result (\ref{EQN_LL}) follows from scaling of luminosity in Poynting flux $L = k_1 \Phi^2$, where $\Phi$ denotes the poloidal magnetic flux
advected to the ISCO and $k_1$ represents remaining parameters such as angular velocity in
the underlying Faraday induction process. In the presence of, e.g., violent instabilities in the inner accretion disk or torus, 
the mass $m(t)$ about the ISCO satisfies $m(t)=\alpha t$ $(0\le t \le \tau)$ in the on-state and $m(t)=0$ ($\tau < t < T)$ in the off-state. 
Here, $\tau/T$ is the duty cycle and $\alpha$ is the instantaneous mass accretion rate onto the ISCO. In advecting $\Phi$
by $m$, $\Phi=k_2 m$ for some constant $k_2$, whereby $L=km^2$ with $k=k_1k_2$. Here, $k$ 
will be effectively constant over intermittent time scales $T$ much less than the lifetime of the source. Accordingly, the mean mass about the ISCO satisfies 
\begin{eqnarray}
<m> = \frac{1}{T} \int_0^\tau \alpha t dt = \frac{\alpha\tau^2}{2T}
\label{EQN_E1}
\end{eqnarray}
with associated mean luminosity
\begin{eqnarray}
<L_w^i>= \frac{k}{T} \int_0^\tau m^2 dt = \frac{k\alpha^2 \tau^3}{3T} = \frac{4}{3} k<m>^2 \frac{T}{\tau}.
\label{EQN_E2}
\end{eqnarray}
Identifying $L_w^c = k <m>^2$ with the luminosity that would be produced by the mean mass at the ISCO, (\ref{EQN_LL}) follows. 

To exemplify, the micro-quasar GRS 1915+105 has $T/\tau$ of order 60. At this (inverse) duty-cycle, (\ref{EQN_LL}) shows 
that intermittencies can enhance the power output by well over an order of magnitude over that of a steady state source at the same
accretion rate. It is tempting to speculate
that at least some of the anomalously large luminosity in its supermassive counterparts such as the quasar
3C273 can similarly be attributed to an appreciable duty cycle, apparent in the ejection of relativistic blobs
at a minimum Lorentz factor $\Gamma = \sqrt{1+\beta_\perp^2}\simeq 10$, where $\beta_\perp\simeq 9.6$
denotes their apparent velocities on the celestial sphere. The same such intermittencies also appear to be advantageous to 
emitting Ultra-High Energy Cosmic Rays from on-average low-luminosity AGN \citep{far09,van09}.

In the propagation of Alfv\'en waves downstream and outside of light cylinder of a central engine, open magnetic winds inevitably become toroidal \citep{con95}. Simulations demonstrate that toroidal magnetization gives rise to the aforementioned distinctive nose-cone, by which the head of the jet propagates relatively fast compared to unmagnetized jets \citep{cla86,van96}. In three-dimensions, simulations show that toroidally magnetized jets are not threatened by the kink instability \citep{mck09}. Magnetized toroidally magnetized outflows hereby appear to be a suitable starting point to address the ubiquity of jet-driven supernovae from intermediate to relatively massive stellar progenitors, and principally different from relativistic hydrodynamics.

Figs. 1 and 2 show the morphological evolution, propagation and breakout from a remnant stellar envelope of ejecta from burst outflows employing a fully four-covariant hyperbolic system of Maxwell's equations in divergence form \citep{van91} in the approximation of axisymmetry.

Fig. 1 shows the formation of a relativistic ejecta following a burst of low density outflow into a stellar interior of relatively high density, modeled by an aperture with time-dependent boundary conditions. Out-of-equilibrium radial pressure due to magnetic pinch at launch produces a characteristic magnetic ring vortex (MTV)-nose cone morphology. The Mach disk in the latter is generally unsteady which tends to shed internal sonic nozzles in its wake, that appear as knotted structures. The results of the burst outflow, in distinction to continuous outflow, is a ballistically propagating nose cone, after the MTV dissolved by an expansion wave. At moderate Lorentz factors, the expansion wave following switch off can even appear as back flow to the source. 

The numerical simulation demonstrates the formation ejecta disconnected from the source, that live long after switch off of the engine. They ultimately escape through a successful breakout into the surrounding region of relatively lower density. In our model, the velocity of sound in the corona just outside the star is larger than inside, giving rise to a distinguishable relativistic expansion in the stellar corona at breakout of the nose-cone. Any high energy emission from the breakout process hereby may derive from both the relativistic Mach disk inside the head as well as from this large area and wide angle relativistic bow shock. 

The formation of a nose-cone as stable ejecta is unique to magnetized outflows, not found in hydrodynamical simulations. 
Switching off a hydrodynamical jet results in dissolving the HTV at the head, leaving only a large angle bow shock that may produce a non-relativistic supernova with no high energy emissions at breakout \citep{laz12}.
Relevant to breakout of magnetized ejecta, therefore, is their total energy rather than the lifetime of the inner engine.  
These ejecta hereby provide a robust explosion mechanism that is remarkably insensitive to the details of the source.

The aspherical CC-SNe with LGRBs and low luminosity GRBs (LLGRBs) such as GRB 980425/SN1998bw \citep{gal98} are 
commonly attributed to the breakout of the bow shock ahead of a relativistic wind or jet. Relativistic shocks at breakout have been associated with the smooth light curves of LLGRBs, e.g., GRB 980425/SN1998bw, GRB060218/SN1006aj \citep{cam06} and GRB080109/SN2008D \citep{sod08,maz08}, here attributed to the breakout of moderately relativistic magnetized ejecta from the inner accretion disk around a rotating black hole. Disk winds and their ejecta hereby provide an efficient aspherical explosion mechanism with and without association to GRBs \citep{mae06,tau09}, that may be especially efficient when baryon rich \citep{van11}.  

In extreme cases, outflows disks may collimate an ultra-relativistic baryon-poor jet from a rapidly rotating black hole within. Ultra-relativistic baryon-poor outflows provide just the kind of agent needed to successfully produce a normal long GRBs. Their prompt GRB emission is believed to result from internal shocks, converting kinetic energy into non-thermal radiation possibly augmented by dissipation of magnetic fields. Intermittency at the source then appears necessary for producing strong internal shocks downstream, here enhanced by intermittency following (\ref{EQN_LL}). Long duration GRB-supernovae hereby result from rare instances of two component outflows, from an ultra-relativistic baryon-poor inner jet within baryon-rich winds or ejecta powering the supernova (e.g. \cite{van03}). 

For intermittencies derived from random processes, we next turn to two outbursts with toroidal field reversal. 

Fig. 2 shows the interaction of two ejecta and the subsequent partial mixing of their toroidal magnetic fields, as the second gradually overtakes and mergers with the first. This mixing is advantageous towards reconnection of the toroidal magnetic field (e.g. \cite{lyu03}), which is otherwise hard to attain in planar shocks rom otherwise similarly intermittent sources \citep{lev97}.

Although this experiment is shown for moderately relativistic ejecta appropriate for baryon-rich disk winds, we advance
the notion of magnetic reconnection by mixing in the merger of relativistic ejecta with random orientation of magnetic fields at work also in ultra-relativistic outflows. Orientation reversal in magnetic fields then results from aforementioned turbulent inner disk or torus around a rotating black hole. Since normal LGRBs require a two-component jet, while LLGRBs may already result from a baryon-rich disk wind, the former is naturally more rare than the second. The associated stellar explosion may result from both. This explains the robustness of CC-SNe and may explain the different branching ratios into LLGRBs and normal LGRBs. 

By (1), Figs. 1 and 2, we propose a paradigm shift for the most extreme sources to be inherently burst-like. Their luminosity is enhanced by intermittency at the source by the inverse of the duty cycle, that gives rise to ballistic ejecta. Merging of these ejecta with randomly oriented magnetic fields leads to naturally to mixing and shocks. Applied to ultra-relativistic outflows, it provides a pathway to a broadband spectrum  \citep{vur13,bel13,gia12,lev12,ker14} from magnetic reconnection \citep{dre02,gia05,lyu03,mck12} and shock acceleration of, respectively, MeV to GeV photons \citep{ste99b}, provided that the reconnection rate is sufficiently high at sufficient entropy \citep{lyu05,lyu10}. 
A detailed analysis thereof falls outside the scope of the present numerical demonstration on the complex geometry of merging ejecta in RMHD.

Our mechanism for creating extremely luminous outflows seems particularly applicable to the most powerful normal LGRBs associated with hyper-energetic supernovae such as GRB 031203/ SN2003lw \citep{mal04} and GRB 030329/SN2003dh \citep{sta03,hjo03,mat03}. 
Without an ultra-relativistic inner jet, the relativistic expansion of the outer layers of the stellar remnant envelope at breakout of 
the moderately relativistic ejecta of Figs. 1 and 2  may account for the relatively more numerous LLGRBs \citep{kul98}, closely 
related to the original proposal of \cite{col68,col70,col74}.

Normal long GRBs are distinct from aforementioned less extreme {\em Swift} class of SGRBEE and LGRBNs exemplified by GRB 050724 \citep{ber05,bar07} and, respectively, GRB 060614 \citep{del06,gal06,fyn06}. The latter may originate in mergers, e.g., of two neutron stars, producing low mass black holes with rapid spin \citep{bai08}. The distinct $E_{iso}$ from normal LGRBs on the one hand and SGRBEE and LGRBNs on the other hand may result from, respectively, hyper-accretion in CC-SNe and limited accretion in mergers: hyper-accretion onto the ISCO may de-stabilize the torus and the inner torus magnetosphere, pushing it to intermittency and hence 
enhanced luminosities by (\ref{EQN_LL}). 

Our RMHD simulations show several distinctive features over hydrodynamical jets: (a) enhanced
luminosity by intermittency; (b) intermittency producing long-lived magnetized ejecta (Fig. 1) that may power relativistic 
aspherical explosions without requiring a long-lived continuous inner engine; (c) intermittency from turbulent engines
producing merging ejecta with randomly oriented magnetic fields (Fig. 2), whose 
reconnection would provide additional energy to high energy emissions in internal shocks. 

Latent accretion on the remnant inner engine has recently been considered to explain flaring, prevalent
post long GRB emission \citep{chi10}, possibly from fall-back of matter from a outer envelope of the progenitor star
at reduced accretion rates (e.g. $\dot{m}(t)\propto t^{-3}$ in \cite{kum08b}). 
In our model, this accretion tail falls onto slowly rotating black holes following spindown. With no feedback, the black hole luminosity is expected 
to scale linearly in $\dot{m}$ at low efficiency (\cite{mck05}). Inhomogeneous accretion flows due to cooling
(e.g. \cite{gam01,ric05,mej05}) hereby may produce low luminosity flaring for an extended period of time.

As energetic and catastrophic events involving rapidly rotating high density matter, aspherical supernovae are candidate sources of gravitational waves. This output would extend for the lifetime of the inner engine, i.e., long from normal GRB-SNe and possibly short or long from LLGRBs. Due to its proximity, the inner disk of SgrA* might similarly be a persistent source of gravitational radiation of interest to future space based gravitational wave detectors. 

\begin{figure*}
\centerline{\includegraphics[width=85mm,height=75mm]{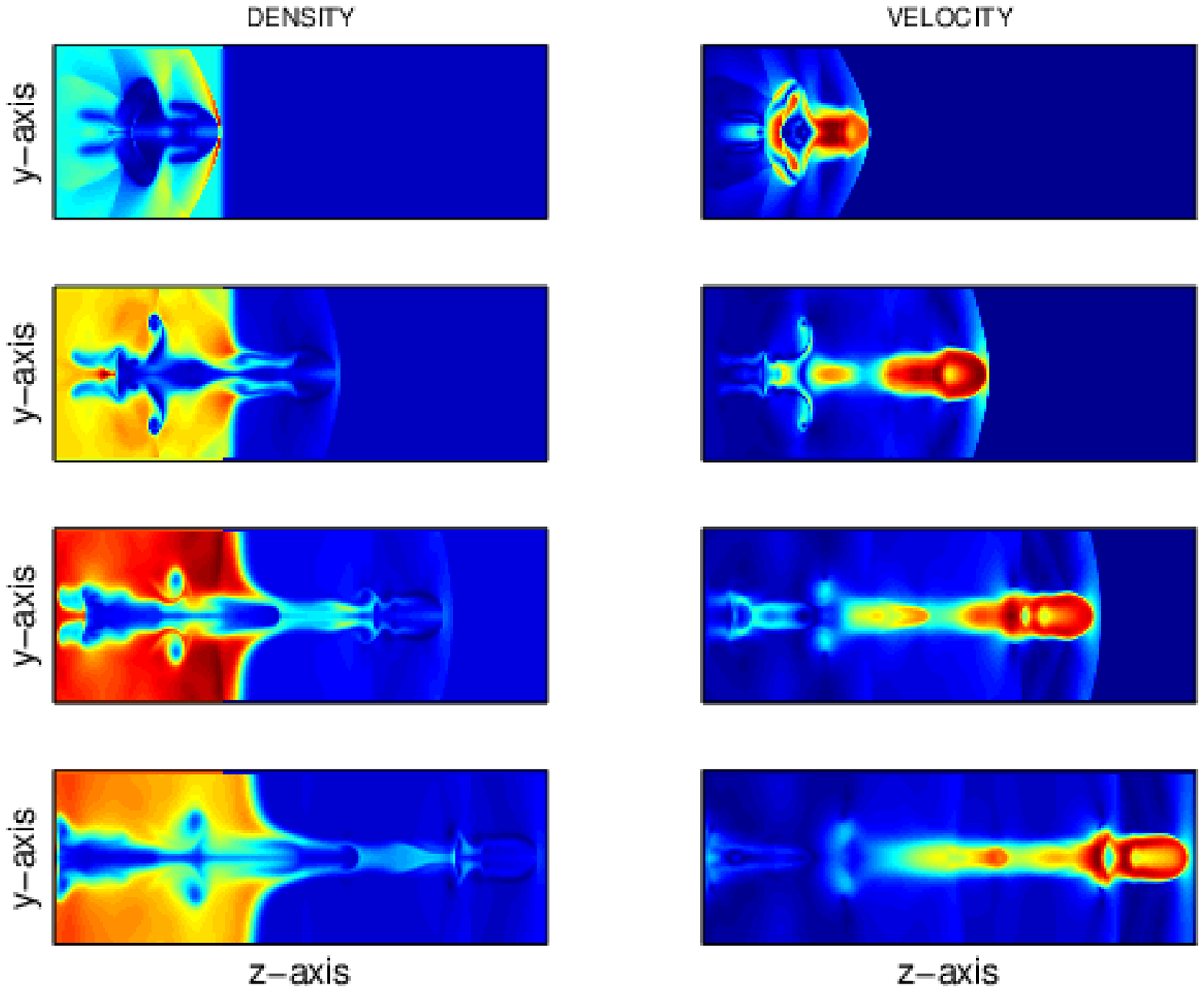}\includegraphics[width=85mm,height=75mm]{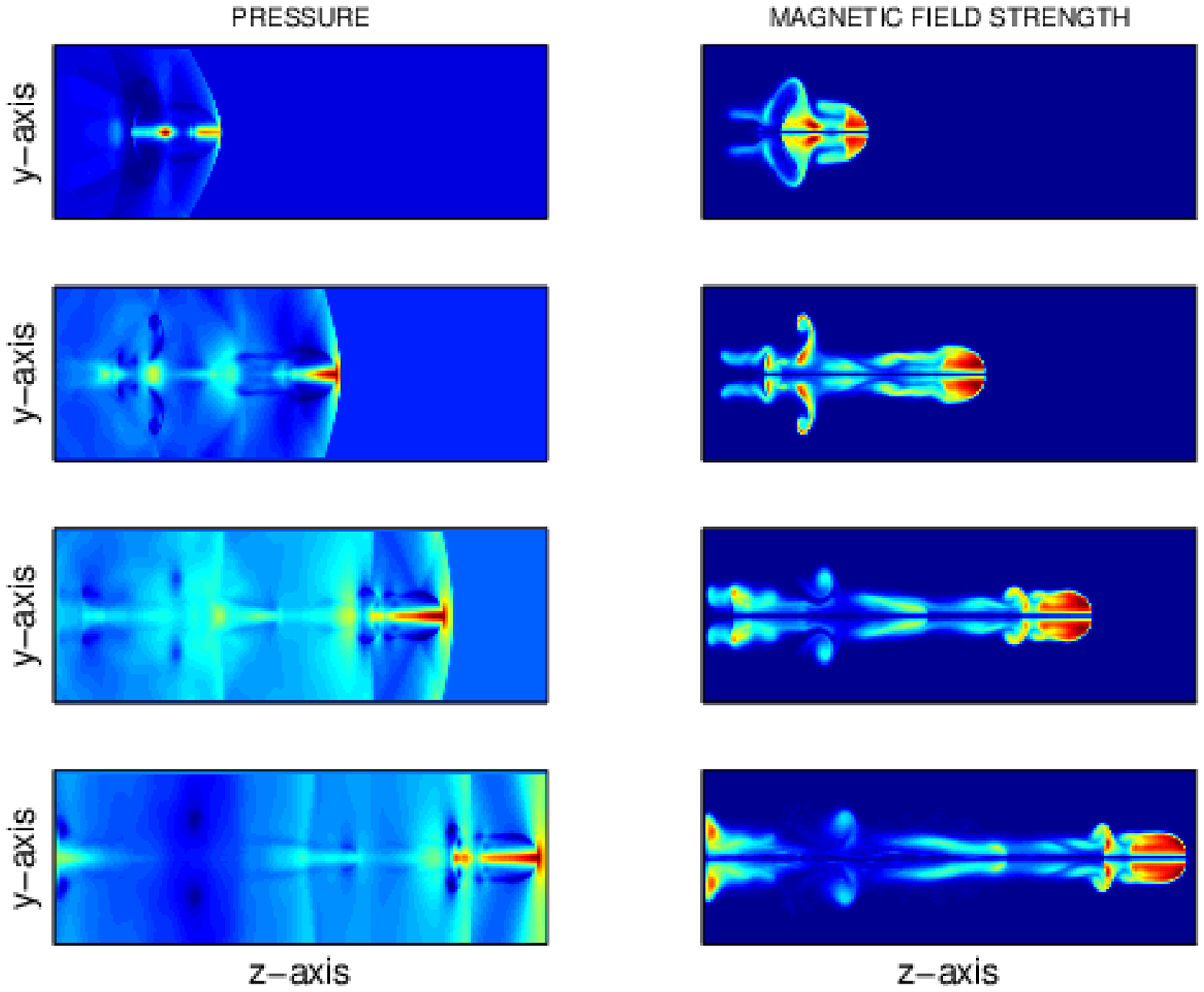}}
\caption{Breakout of relativistic magnetized ejecta produced by a burst with out-of-radial force balance boundary conditions at a source (left) at four consecutive times (top to bottom rows). Shown are distributions of rest mass density, velocity, pressure and magnetic field strength (left to right columns). The velocity and pressure distributions show a magnetized ring vortex (MRV) dissolving soon after switch off, following by a back flow towards the source and a disconnect of the nose-cone, that continues to propagate ballistically and preserving its shape when moving into the low density stellar corona. False colors vary linearly from blue to red with the quantity shown. At the aperture on the left, $(\Gamma, H, P, r) = (2.46, 0.46, 0.10, 0.20)$ in the on-state of the source and $(0,0,0.10,0.20)$ in the off-state.
(Ancillary movie clips series anim1*.mp4 for velocity (V), density (r), magnetic field strength (B) and pressure (P).)
}
\label{fig:A2}
\end{figure*}

\begin{figure*}
\centerline{\includegraphics[width=85mm,height=75mm]{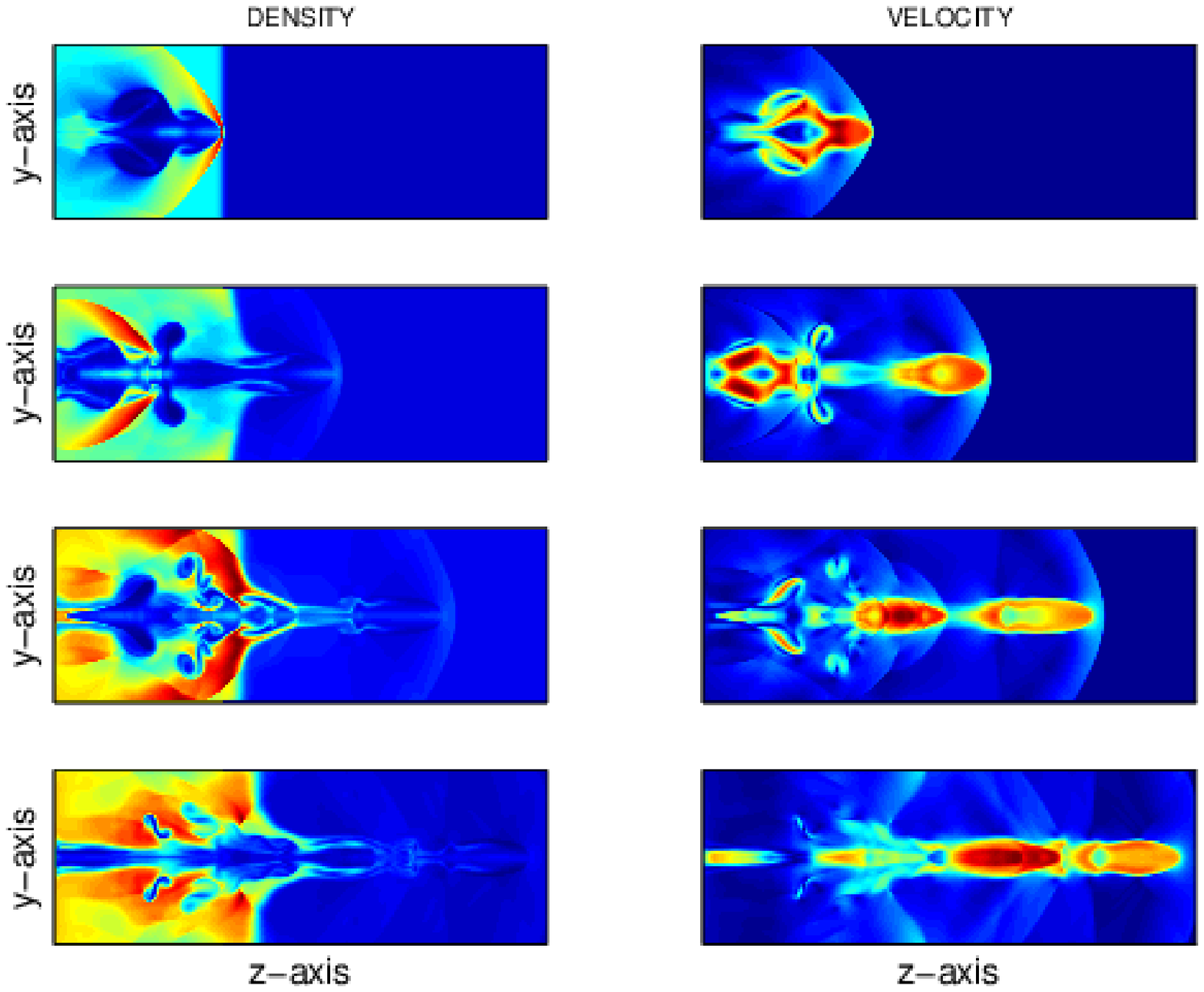}\includegraphics[width=85mm,height=75mm]{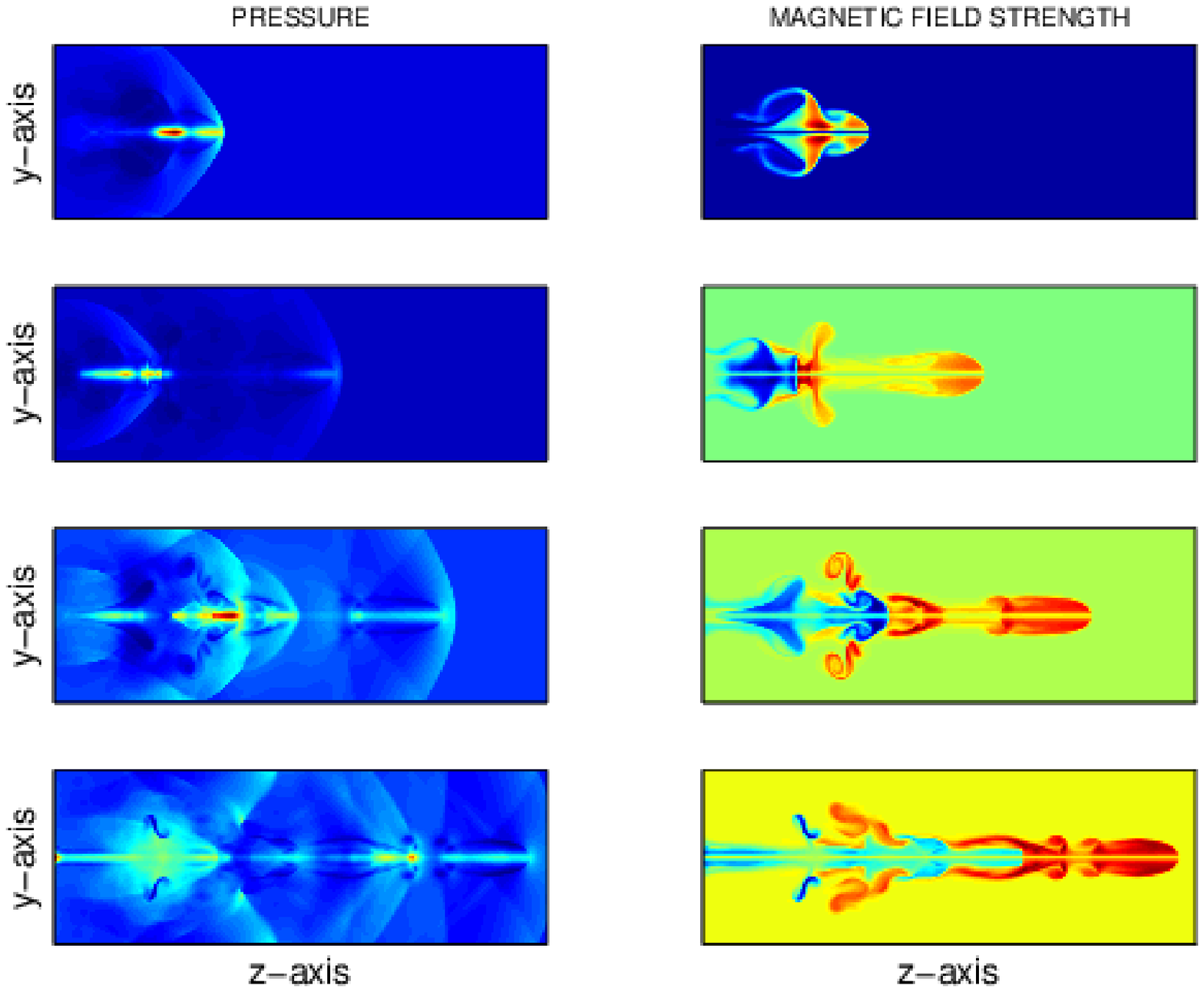}}
\caption{Evolution of two ejecta following Fig. 1 with have opposite orientation of the toroidal magnetic (largely blue and red in the column to the right). As the second catches up and penetrates the first, oppositely oriented magnetic fields are expected to mix and reconnect. At the aperture $(\Gamma, H, P, r) = (2.15, 0.40, 0.10, 0.20)$ in the on-state of the source and $(0,0,0.10,0.20)$ in the off-state.
(Ancillary movie clips series anim2*.mp4.)
}
\label{fig:A2}
\end{figure*}

{\bf Acknowledgments.} The author gratefully acknowledges A. Levinson, S. Antonucci, M. Della Valle, R. Narayan and the anonymous referee for constructive comments.

\end{document}